\documentclass[aps,prl,reprint,amsmath,amssymb,superscriptaddress,nofootinbib,noeprint]{revtex4-2}
\pdfoutput=1

\usepackage{graphicx}% Include figure files
\usepackage{dcolumn}% Align table columns on decimal point
\usepackage{bm}% bold math
\usepackage[version=3]{mhchem}
\usepackage{mathtools}
\PassOptionsToPackage{hyphens}{url}
\usepackage{float}
\usepackage{color}
\usepackage{graphicx}
\usepackage{amsmath,amssymb}
\usepackage{upgreek}
\usepackage{psfrag} 
\usepackage{graphicx}
\usepackage{braket}
\usepackage{units}
\usepackage{times}
\usepackage[svgnames]{xcolor}
\definecolor{myColor}{rgb}{0.02,0.12,0.3}
\definecolor{myciteColor}{rgb}{0.39,0.7,0.89}
\usepackage[colorlinks=true,citecolor=myColor,linkcolor=myColor,urlcolor=myColor]{hyperref}

\newcommand{\RNum}[1]{\uppercase\expandafter{\romannumeral #1\relax.}}
\makeatletter
\def\maketitle{
\@author@finish
\title@column\titleblock@produce
\suppressfloats[t]}
\begin{document}

\title{On the Stability of the Repulsive Fermi Gas with Contact Interactions}

\author{Yunpeng Ji$^\ast$}
\affiliation{Department of Physics, Yale University, New Haven, Connecticut 06520, USA}
\author{Grant L. Schumacher}
\affiliation{Department of Physics, Yale University, New Haven, Connecticut 06520, USA}
\author{Gabriel G. T. Assumpção}
\affiliation{Department of Physics, Yale University, New Haven, Connecticut 06520, USA}
\author{Jianyi Chen}
\affiliation{Department of Physics, Yale University, New Haven, Connecticut 06520, USA}
\author{Jere Mäkinen}
\affiliation{Department of Physics, Yale University, New Haven, Connecticut 06520, USA}
\affiliation{Yale Quantum Institute, Yale University, New Haven, Connecticut 06520, USA}
\author{Franklin J. Vivanco}
\affiliation{Department of Physics, Yale University, New Haven, Connecticut 06520, USA}
\author{Nir Navon}
\affiliation{Department of Physics, Yale University, New Haven, Connecticut 06520, USA}
\affiliation{Yale Quantum Institute, Yale University, New Haven, Connecticut 06520, USA}


\date{\today}

\begin{abstract}
We report the creation and the study of the stability of a repulsive quasi-homogeneous spin-$1/2$ Fermi gas with contact interactions. For the range of scattering lengths $a$ explored, the dominant mechanism of decay is a universal three-body recombination towards a Feshbach bound state. We observe that the recombination coefficient $K_3\propto \epsilon_\text{kin} a^6$, where the first factor, the average kinetic energy per particle $\epsilon_\text{kin}$, arises from a three-body threshold law, and the second one from the universality of recombination.
Both scaling laws are consequences of Pauli blocking effects in three-body collisions involving two identical fermions. As a result of the interplay between Fermi statistics and the momentum dependence of the recombination process, the system exhibits non-trivial temperature dynamics during recombination, alternatively heating or cooling depending on its initial quantum degeneracy. The measurement of $K_3$ provides an upper bound for the interaction strength achievable in equilibrium for a uniform repulsive Fermi gas.
\end{abstract}

\maketitle

%\section{}

Repulsive interactions in Fermi systems are at the heart of some of the most interesting phenomena in quantum many-body physics. For instance, the interplay between the spin and orbital degrees of freedom gives rise to Stoner's itinerant ferromagnetism in the continuum~\cite{Stoner_1933} and to the complex phases of the repulsive Hubbard model on a lattice~\cite{mielke_1993}.

The dilute repulsive spin-1/2 Fermi gas, where the interactions between two spin states $\uparrow$ and $\downarrow$ are described by a positive $s$-wave scattering length $a$, is one of the most fundamental quantum many-body models~\cite{Huang_1957,lee_1957,Galitskii_1958}. Among its important features, it is amenable to first-principle calculations in perturbation (for $k_\text{F}a\ll 1$, where $k_\text{F}$ is the Fermi wavenumber). In that limit, its properties (e.g. ground-state energy, Landau parameters, etc.) are universal, i.e. they depend on $a$ alone, not on details of short-range physics~\cite{Galitskii_1958,Landau_1957,Efremov_2000}.

Ultracold atomic gases have emerged as a powerful platform for studying this model, because effective repulsion can be implemented on the so-called `upper' (repulsive) branch using short-range attractive potentials~\cite{Jo_2009,Sanner_2012,Lee_2012,Valtolina_2017,Scazza_2017,Amico_2018,Scazza_2020}. This implementation is particularly interesting because it can realize the regime of strong ($k_\text{F}a\gtrsim 1$) yet short-range interactions ($k_\text{F}r_0\ll 1$, where $r_0$ is the potential range), see e.g.~\cite{Pricoupenko_2004,Shenoy_2011}.

However, the repulsive Fermi gas with short-range attractive potentials is intrinsically metastable. This originates from the existence of a universal bound state in the two-body problem for $a>0$, with a binding energy $\epsilon_\text{b} = \frac{\hbar^2}{m a^2}$ where $m$ is the mass of the atom. The pairing instability of the repulsive branch of the many-body system towards the lower (attractive) branch of bound pairs, depicted in Fig.~\ref{FIG:1}(a), is a complex problem; it is expected to evolve from an instability driven by \emph{universal} three-body recombination for $\epsilon_\text{b}\gg E_\text{F}$~\cite{Petrov_2003,Esry_2001}, to many-body pairing effects when $\epsilon_\text{b}\lesssim E_\text{F}$~\cite{Petrov_2003,Pekker_2011,He_2016,Amico_2018} where $E_\text{F}$ is the Fermi energy.

This pairing instability has played a central role in the study of the strongly repulsive Fermi gas and the search for the itinerant-ferromagnet phase~\cite{Duine_2005,LeBlanc_2009,conduit_2009,conduit_2009_2,conduit_2009_3,chang_2010,schmidt_2011,pilati_2010,von_2011,chang_2011,Shenoy_2011,massignan_2013,pilati_2014,zintchenko_2016,He_2016}. Pioneering experiments have shown decreased lifetime of the gas with increasing interactions~\cite{Jo_2009,Sanner_2012} and larger initial rate of reduction of repulsive correlations (possibly due to the ferromagnetic instability) compared to the initial pairing rate~\cite{Amico_2018,Scazza_2020}. 

However, complex dynamics arising from the in-trap density inhomogeneity as well as the far-from-equilibrium nature of the initial quenched states have hindered the study of the homogeneous system's stability~\cite{Jo_2009,Amico_2018}. The advent of homogeneous gases prepared in optical box traps~\cite{Gaunt_2013,Chomaz_2015,Mukherjee_2017,navon_2021} has enabled the investigation of complex stability problems in clean settings~\cite{eigen_2017,bause_2021,shkedrov_2022}. Here, we revisit the fundamental problem of the stability of the repulsive Fermi gas by measuring the three-body recombination law in a homogeneous gas. 

The experiment starts with a weakly attractive gas of $^6$Li atoms in a balanced mixture of the first and third lowest Zeeman sublevels (respectively labeled as $\uparrow$ and $\downarrow$), trapped in a red-detuned optical dipole trap. The gas is evaporatively cooled at a bias magnetic field $B= 287$~G. It is then loaded in a blue-detuned (at a wavelength of $639$~nm) cylindrical box trap constructed by intersecting a `tube' beam (produced with a set of axicons) with two thin sheets, see Fig.~\ref{FIG:1}(b). The magnetic field is then ramped to $B=597$~G where the interactions are weakly repulsive ($a \approx 500~a_0$, where $a_0$ is the Bohr radius ~\cite{zurn_2013_2}). At this stage, we typically have $N_{\uparrow} \approx N_{\downarrow} \approx 6 \times 10^5$ atoms per spin state at $T \approx 0.3~T_\text{F}$ with $E_\text{F} \approx k_{\mathrm{B}} \times 0.5\;\mu\mathrm{K}$ and a spin imbalance of $\frac{N_{\downarrow}-N_{\uparrow}}{N_{\downarrow} + N_{\uparrow}} = 0.2(3)\%$. The interaction field is then ramped to its final value over $100$~ms, and left to settle for an additional $25$~ms. We then hold the atoms for a variable duration $t_\text{hold}$. We image the gas near the zero crossing of $a$ ($|a| \le 50~a_0$) by quickly ramping the field to $B= 569$~G, so that trapped pairs are converted into tightly bound molecules and thus detuned from the atomic imaging resonance~\cite{imaging,SuppMat}.

\begin{figure}[!h]
\includegraphics[width=1\columnwidth]{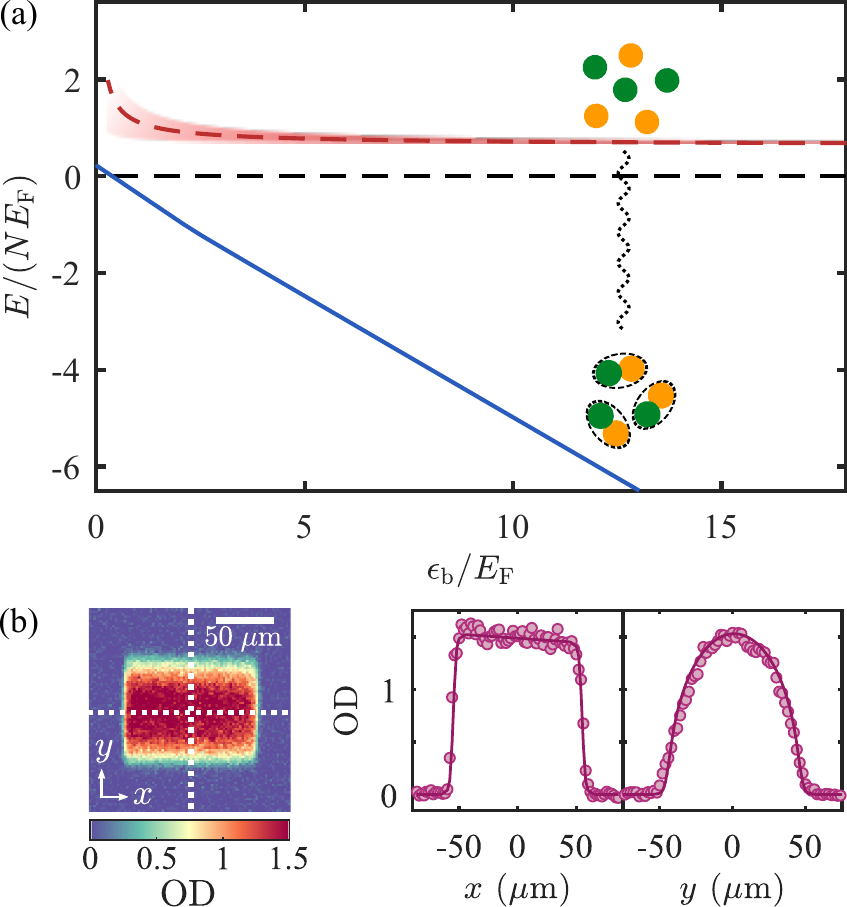}
\caption{A homogeneous repulsive Fermi gas prepared in an optical box. (a) Sketch of the two lowest energy branches of a Fermi gas with a positive scattering length $a$; the `upper' (repulsive) branch is shown in red, the `lower' branch (a gas of fermion pairs) is shown in blue. The red dashed line is the repulsive Fermi gas energy up to second order in $k_\text{F}a$~\cite{Huang_1957,lee_1957}; the red shaded area depicts the energy width associated with the finite lifetime of the upper branch. (b) \emph{In-situ} imaging of the box-trapped Fermi gas. Gravity, here oriented along $-\mathbf{\hat{y}}$, is compensated by magnetic levitation. The image on the left is the column-integrated optical density (OD). The plots on the right are cuts along the white dashed lines of the image. The solid lines are derived from the fit used to extract the volume of the box; $V=7.3(6) \times 10^{-4}~\mathrm{mm}^3$. The slanted profile in the horizontal cut is caused by the slightly conical shape of our cylindrical box~\cite{SuppMat}.}
\label{FIG:1}
\end{figure}
We show in Fig.~\ref{FIG:2}(a) examples of time evolution of the atom number $N$ per spin state for different values of $a$, normalized to the initial atom number $N_0$. Qualitatively, the gas lifetime decreases with increasing $a$, even though $N_0$ also decreases (because of losses during the interaction field ramp and the settling time~\cite{SuppMat}). The average kinetic energy per particle $\epsilon_\text{kin}$, measured after time-of-flight expansion and shown in Fig.~\ref{FIG:2}(b), also slowly decreases with $t_\text{hold}$.

The origin of the decay is model-independently revealed by plotting the atom loss rate $\dot{N}/N_0$ versus $N/N_0$ (Fig.~\ref{FIG:2}(c)). The examples shown follow a scaling relation of the rate $\dot{N} \propto -N^{\gamma}$ (fits are shown as solid lines, and fitted values of $\gamma$ are in legend). We observe that $\gamma\approx 1$ at weak interactions ($a\ll 10^3~a_0$) where the losses are caused by density-independent collisions with the residual background gas. For stronger interactions, we observe $\gamma\approx 3$, consistent with an atom loss rate per unit volume
\begin{equation}\label{eq:loss}
\dot{n}  =  -L_3 n^3
\end{equation}
due to three-body collisions, with a constant loss coefficient $L_3$ and a uniform density $n=N/V$, where $V$ is the volume of the box.

\begin{figure}[!h]
\includegraphics[width=1\columnwidth]{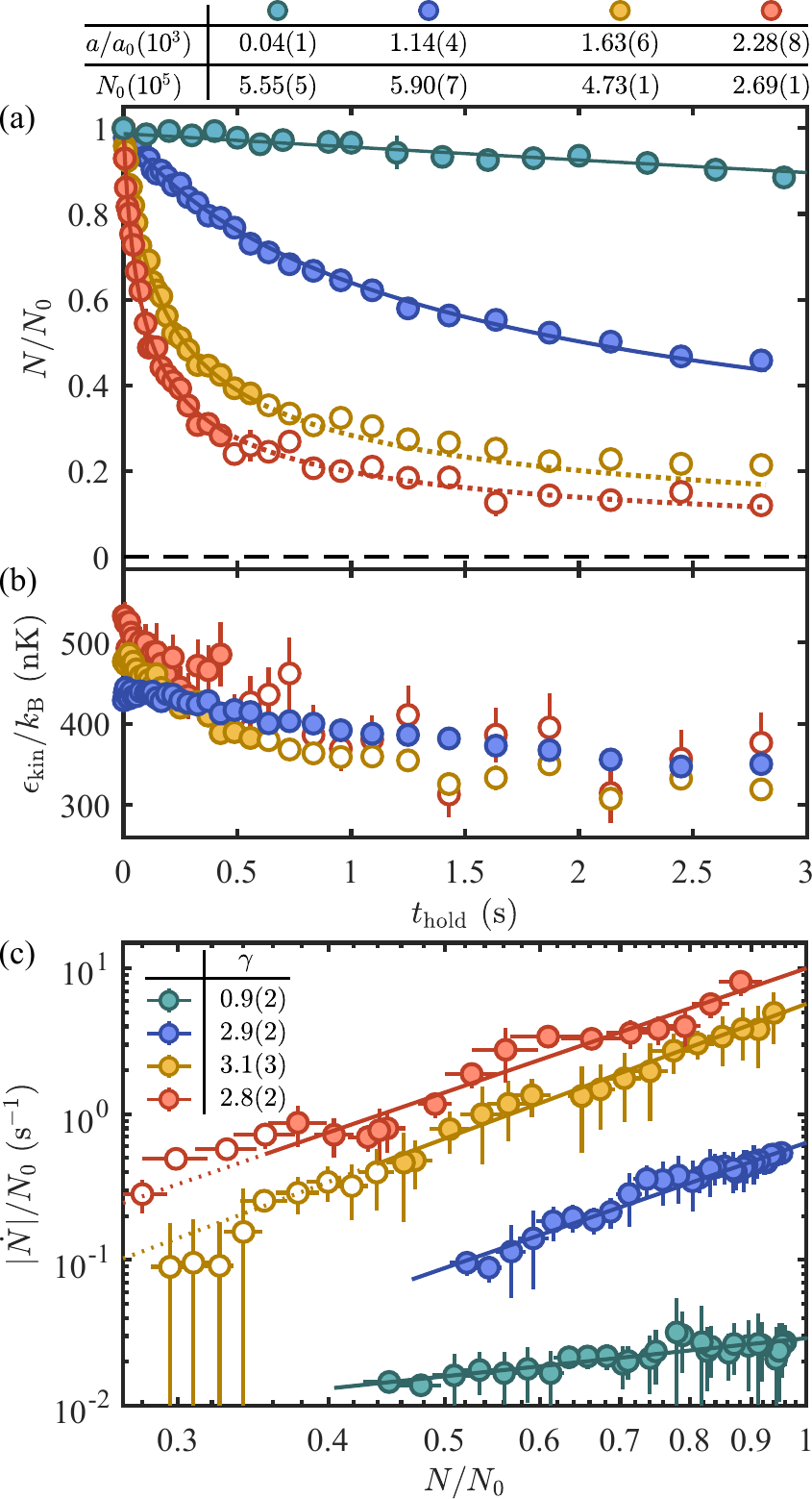}
\caption{Decay of a uniform repulsive Fermi gas. (a) Evolution of atom numbers for different interaction strengths, normalized to the initial atom numbers $N_0$. The solid blue, yellow, and red lines are fits to a three-body loss model that includes a one-body loss rate determined from the green-line fit~\cite{onebody}. The three-body loss fits are limited to the region where $\epsilon_\text{kin}$ changes by less than $20\%$ of its initial value, indicated by solid circles; open circles are not used in the fit. The same marker style is used in (b) and (c). Dotted lines are extensions of the fits beyond the fitting range. (b) Evolution of the average kinetic energy per particle during atom losses. (c) Scaling relation between atom loss rate and atom number. Solid lines are power law fits and the extracted exponents $\gamma$ are listed in the legend. }
\label{FIG:2}
\end{figure}

\begin{figure*}[!hbt]
\centerline{\includegraphics[width=\textwidth]{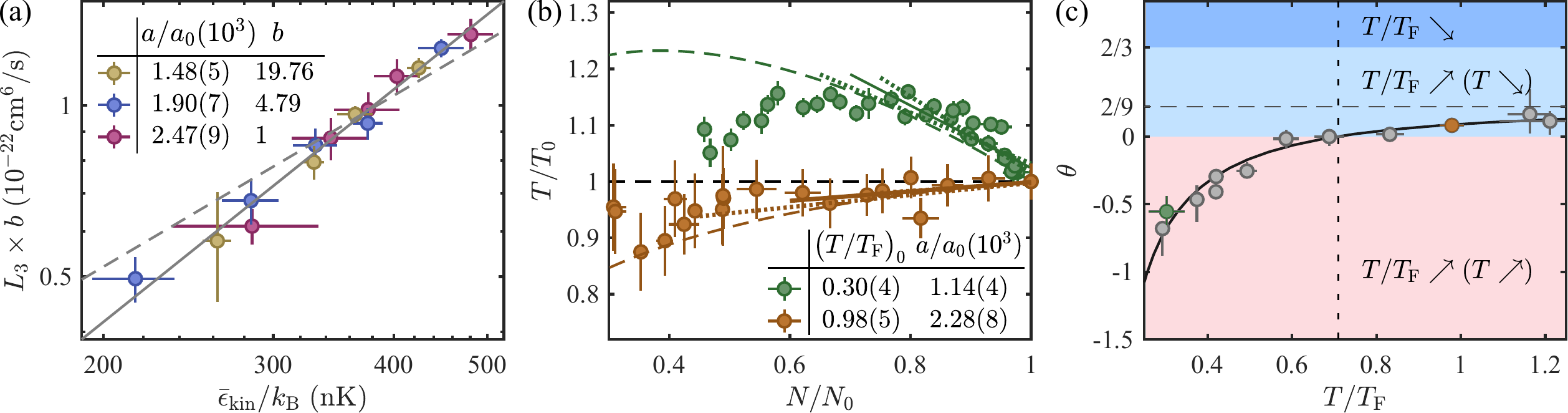}}
\caption{Threshold law of three-fermion recombination. (a) Scaling relation between $L_3$ and the (time-averaged) kinetic energy $\bar{\epsilon}_\text{kin}$. The solid line shows the power law fit on three sets of data, which is rescaled by a factor for clarity (see legend). The dashed line is the fit assuming $\lambda = 1$~\cite{SuppMat}. (b) Temperature evolution during three-body losses. The dashed lines are theoretical predictions without adjustable parameters, given the initial measured $(T/T_\text{F})_0$ (see legend). The solid lines are linear fits to extract the coefficient $\theta$; the dotted lines show estimate on the uncertainty of $\theta$, see panel (c). (c) Temperature-change coefficient $\theta$ versus $T/T_\text{F}$. The vertical dashed line marks the critical $(T/T_\text{F})^*$ at which $\theta$ changes sign, and the horizontal dashed line shows the asymptotic value of $\theta$ in the classical limit.}
\label{FIG:3}
\end{figure*}

Now that we have established a range over which losses are dominated by three-body recombination, we quantitatively characterize the process. The event rate per unit volume for each type of event is $ \Omega \equiv K_3 n^3$ ($=\Omega_{\uparrow\uparrow\downarrow} = \Omega_{\uparrow\downarrow\downarrow}$) where $K_3$ is the recombination coefficient; $K_3$ can be studied through losses, since $K_3 =  L_3/d$, where $d$ is the average number of atoms lost per event (either because their release energy from recombination exceeds the trap depth or because they form molecules that are optically detuned). We obtain $L_3$ by fitting $N(t)$ to the solution of Eq.~(\ref{eq:loss})~\cite{onebody} (solid lines in Fig.~\ref{FIG:2}(a)). To ensure that $L_3$ is approximately constant with $t_\text{hold}$, the fits are restricted to a range where ${\epsilon}_\text{kin}$ changes by at most $20\%$ of the initial value, see solid points in Fig.~\ref{FIG:2}~\cite{SuppMat}.

We examine this assumption more carefully by studying the relationship between $L_3$ and $\epsilon_\text{kin}$. We control $\epsilon_{\text{kin}}$ by varying the box depth at an intermediate evaporative cooling stage, keeping the final box depth $U_\text{box}$ the same. As shown in Fig.~\ref{FIG:3}(a) for three different values of $a$, we observe that $L_3$ scales as a power law of $\epsilon_\text{kin}$ averaged over time, $\bar\epsilon_\text{kin}$. 

Theoretically, $K_3\propto \epsilon_\text{kin}^\lambda$, where the exponent $\lambda$ is determined by the three-body threshold laws, which crucially depends on the symmetries imposed by the quantum statistics of the collision participants~\cite{Esry_2001}. For instance, for three distinguishable particles or indistinguishable bosons, there is no energy dependence ($\lambda=0$); for three indistinguishable fermions, $\lambda=2$ ~\cite{Yoshida_2018,top_2021}. The generic process in the spin-$1/2$ Fermi gas corresponds to the previously-unverified case of collisions involving two indistiguishable fermions. The three-body event rate in a unit volume $\omega_3$ depends on the momenta $\mathbf{k}_1$ and $\mathbf{k}_2$ of the indistinguishable fermions, and is independent of the third participant's momentum $\mathbf{k}'$~\cite{petrov}:
\begin{equation}\label{eq:omega}
\omega_3(\mathbf{k}_1,\mathbf{k}_2,\mathbf{k}')\propto (\mathbf{k}_1-\mathbf{k}_2)^2.
\end{equation} 
Integrating Eq.~(\ref{eq:omega}) over the phase space density of the three participants, one finds $\lambda=1$. Experimentally, we measure $\lambda_\text{exp} = 1.36(14)$~\cite{epsilonkin} (solid line in Fig.~\ref{FIG:3}(a)) , in reasonable agreement with the theoretical prediction.

The dependence of $\omega_3$ on momentum has interesting implications on the temperature dynamics of the gas during decay. In Fig.~\ref{FIG:3}(b), we show $T/T_0$ versus $N/N_0$ (where $T_0$ is the initial $T$). Depending on $T/T_\text{F}$, the system either cools down or heats up.  
This effect results from an interplay between Fermi correlations and the momentum dependence of $\omega_3$. The cooling effect from the preferential removal of particles with large momenta (without spatial selectivity)~\cite{SuppMat}, strongest for $T\gg T_\text{F}$, competes with the heating from the perforation of the Fermi sea, which dominates in the deeply degenerate regime~\cite{timmermans_2001}. A theoretical model for a closed system, shown as colored dashed lines in Fig.~\ref{FIG:3}(b), yields good agreement with the observed evolution of the temperature for $N/N_0\gtrsim0.7$~\cite{SuppMat}. We attribute the discrepancy at late times for low $(T/T_\text{F})_0$ to additional cooling from plain evaporation.

Quantitatively, we define the coefficient $\theta \equiv \frac{N}{T}\left(\frac{\partial T}{\partial N}\right)_V$ under this rarefaction, and measure it at $t_{\text{hold}}=0$ for various $T/T_\text{F}$ (Fig.~\ref{FIG:3}(c)). We observe that the transition from heating to cooling occurs at a critical degeneracy $(T/T_\text{F})^* \approx 0.7$. 
The measurements are in excellent agreement with the theoretical prediction (solid line in Fig.~\ref{FIG:3}(c))~\cite{SuppMat}, which establishes the crossing at $(T/T_\text{F})^* = 0.71$ (vertical dashed line). For $T\gg T_\text{F}$, $\theta$ approaches $2/9$, where the cooling effect is most pronounced. Note that for all $T$, $\theta<2/3$, so that this process does not increase the quantum degeneracy of the gas (see related scenarios for bosons~\cite{schemmer_2018,Dogra_2019}, and fermions near a narrow Feshbach resonance~\cite{Peng_2021}). 

We now turn to the dependence of $L_3$ on interactions. In Fig.~\ref{FIG:4}(a), we display $\gamma$ versus $a$; the solid points are data where losses are three-body dominated (see Fig.~\ref{FIG:4} and caption). 
We subsequently extract $L_3$ for all interactions by fixing $\gamma=3$ and taking one-body decay into account~\cite{onebody}; to factor out the effect of the threshold law, we display $L_3/\bar \epsilon_{\text{kin}}$, see Fig.~\ref{FIG:4}(b). We observe that over more than four orders of magnitude, $L_3/\bar \epsilon_{\text{kin}}$ follows a power law of $a$. Fitting the data in the three-body-dominated region (solid blue points in Fig.~\ref{FIG:4}(b)), we find $L_3/\bar \epsilon_{\text{kin}} \propto a^{6.1(2)}$ (solid blue line).

The fact that $L_3$ scales precisely as $a^6$ is strong evidence for the universality of this  process. Indeed, should three-body recombination be universal, i.e. be independent of short-range physics, the threshold law implies the scaling of $K_3$ with interaction strength~\cite{DIncao_2005}. Specifically, if $K_3\propto \epsilon_\text{kin}^\lambda$, then on dimensional grounds one should have $K_3\propto \epsilon_\text{kin}^\lambda \frac{m^{\lambda-1}}{\hbar^{2\lambda-1}}a^{4+2\lambda}$. For two identical fermions, one finds $K_3\propto a^6$, in excellent agreement with our measurements. It is interesting to note that the $a^4$ scaling for bosons is not universal, due to effects related to Efimov physics~\cite{braaten_2007,naidon_2017}.
Compared to the bosonic case, an additional factor $\epsilon_{\text{kin}}/{\epsilon_\text{b}}$, $\propto(k_\text{F}a)^2$ at low $T$, can be interpreted as a suppression factor due to Pauli blocking, which arises as two identical fermions need to come within $\approx a$ of each other to form a final bound state.

Now that we established $L_3 \propto \epsilon_\text{kin} a^6$, we can extract the dimensionless constant $A$ in $L_3 = d A \epsilon_{\text{kin}} a^6/\hbar$, predicted to be universal~\cite{Auniv}. As some or all products of the recombination can be lost, $d$, the link between losses and recombinations, depends on the box depth $U_\text{box}$ and $\epsilon_\text{b}$. To gain insight into this link, we implement a second imaging protocol where we image the atoms directly at the interaction field (depicted in the top left inset of Fig.~\ref{FIG:4}(b)); in our range of $a$, molecules and atoms are optically unresolved~\cite{imaging}. The measurements are displayed as red circles in Figs.~\ref{FIG:4}(a)-(b).

\begin{figure}[!t]
\includegraphics[width=1\columnwidth]{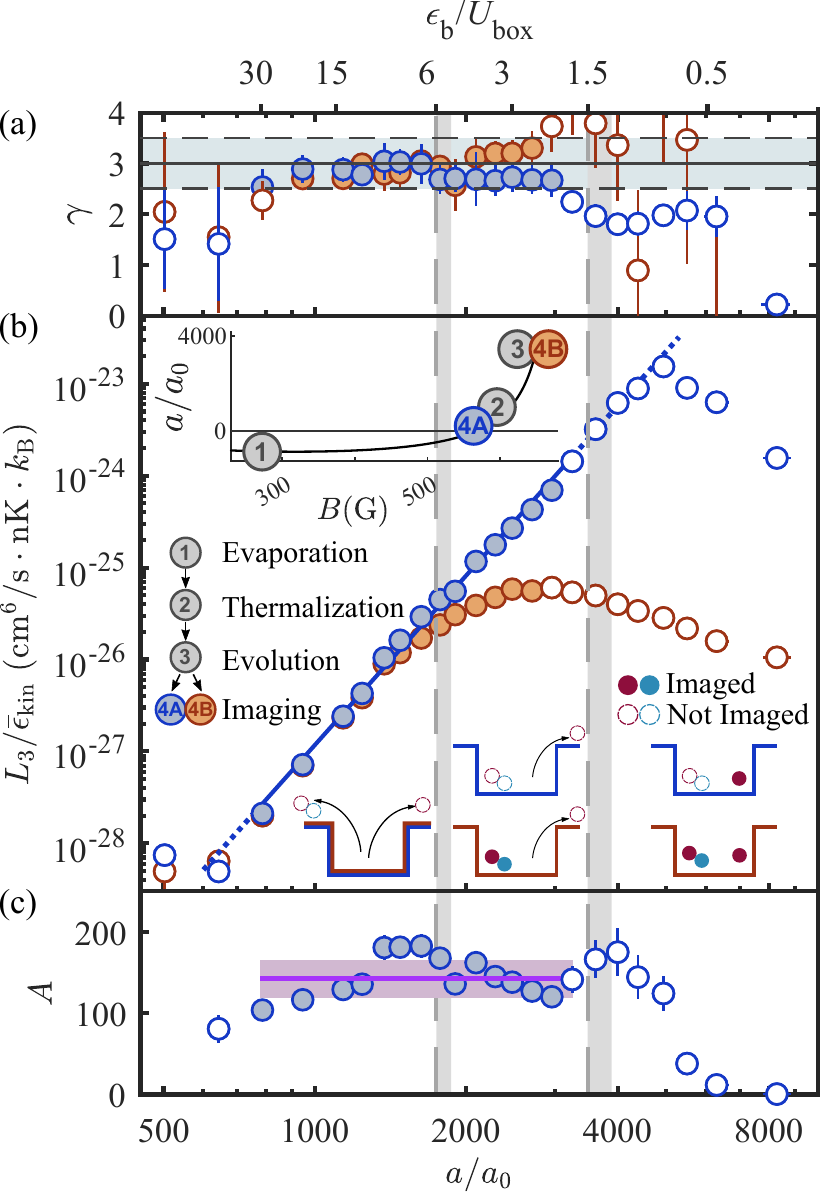}
\caption{Universality of three-body recombination. (a) Atom-loss scaling exponent $\gamma$. Blue and red circles are respectively imaged near the zero crossing of $a$ or directly at the interaction field. Data in the three-body dominant region, selected by $|\gamma - 3| \le 0.5$ (blue band) and with a relative uncertainty $\le 20\%$, are shown by solid points and left open otherwise, in all panels. (b) Universal scaling of $L_3$ with $a$. The experiment sequence is shown in the upper insets. The blue line is the power law fit on the solid blue points. Vertical grey dashed lines mark the threshold values of $a$ such that $\epsilon_\text{b}/3 = 2 U_{\text{box}}$ and $2\epsilon_\text{b}/3 = U_{\text{box}}$, and the bands include average over initial energies ~\cite{SuppMat}. Bottom cartoons depict imaging and trapping regimes after recombinations for the atoms and molecules. (c) Universal constant $A$. Data points are the experimental values of $A = \hbar L_3/(3 \bar\epsilon_{\text{kin}} a^6)$, and the solid purple line is derived from a global $a^6$ fit to the data in (b) (not shown). The systematic error from the volume calibration is shown by the light purple band \cite{SuppMat}.}
\label{FIG:4}
\end{figure}

At low $a$, $L_3$ measured by both imaging methods coincide, as $d=3$ in both cases. The separation at $a \gtrsim 1300~a_0$ occurs close to the condition $\epsilon_\text{b}/3 \approx 2 U_{\text{box}}$ at which the molecules remain trapped (see cartoons at the bottom of Fig.~\ref{FIG:4}(b))~\cite{deposit}. For larger $a$, $d<3$ for the `interaction field' imaging. 

For the `zero-crossing' imaging, $d=3$ still holds; the $a^6$ scaling extends up to the point where $2\epsilon_\text{b}/3 < U_{\text{box}}$, beyond which all recombination products may be trapped~\cite{Petrov_2003,unitary}. The maximum of $L_3(a)$ is located marginally beyond this threshold. Fixing $d=3$, we fit $L_3/\bar \epsilon_{\text{kin}}$ (solid blue points) and find $A = 143(16)_{\mathrm{stat.}}(24)_{\mathrm{sys.}}$. To examine more closely the quality of the $a^6$ scaling, we extract $A$ without free parameters from $(\hbar L_3/(3 \bar\epsilon_{\text{kin}})/a^6$ (Fig.~\ref{FIG:4}(c)). Our measurements are in excellent agreement with the theoretical prediction $A=148$ for the mass-balanced three-fermion problem~\cite{Petrov_2003}. 

The range over which the $a^6$ scaling law applies is surprisingly large. First, it extends even at large $a$ where the measured $\gamma$ is only marginally close to $3$ (see open circles in Fig.~\ref{FIG:4}). Secondly, at the highest $a$ for which we observe $a^6$ scaling, $\epsilon_{\text{kin}}\gtrsim k_\text{B}\times 0.5$~$\mu$K is only slightly smaller than $\epsilon_\text{b}\approx k_\text{B}\times 1.2$~$\mu$K, even though the condition for the universal scaling is expected to be valid for $\epsilon_{\text{kin}} \ll \epsilon_\text{b}$~\cite{Petrov_2003}.

Finally, our measurement of $K_3$ provides an important ingredient for assessing the limits of equilibrium for a strongly interacting repulsive Fermi gas. To ensure equilibrium, $\Gamma_3 \equiv 3 K_3 n^2$~\cite{gamma3} must be significantly slower than $\Gamma_2$, the two-body elastic collision rate. We find $\Gamma_2/\Gamma_3 = (k_{\mathrm{F}} a)^{-4} I(T/T_\mathrm{F})$ where $I(T/T_\mathrm{F})$ is a universal function that reaches its maximum at $T \approx 1.2~T_{\mathrm{F}}$. At this temperature, $\Gamma_2=\Gamma_3$ at $k_\text{F} a \approx 1.3$, providing an upper bound to the interaction strength of a repulsive Fermi gas in equilibrium~\cite{SuppMat,kFalim}. This limit is close to the predicted point for the ferromagnetic transition, $k_\text{F} a=\pi/2$ in the mean-field approximation~\cite{houbiers_1997} and $\approx 1$ in quantum Monte Carlo simulations~\cite{pilati_2010,chang_2011,He_2016}.

In conclusion, we studied the stability of the repulsive Fermi gas with short-range interactions. We measured the universal recombination law for three particles of equal mass involving two identical fermions. This work paves the way for the study of complex stability problems of Fermi systems in clean uniform settings, e.g. multi-component gases~\cite{ottenstein_2008,huckans_2009,nakajima_2010}, mass-imbalanced mixtures~\cite{Taglieber_2008,Wille_2008,Barontini_2009,Pires_2014,Tung_2014}, and molecules~\cite{hoffmann_2018,duda_2022}. A future work could leverage uniform Fermi gases to explore the regime $\epsilon_\text{b}\lesssim\epsilon_{\text{kin}}$, where $K_3\propto \epsilon_{\text{kin}} a^6$ should no longer hold, and at low temperature many-body pairing mechanisms are expected to take over~\cite{Pekker_2011,He_2016}. To access the shorter time scales expected, fast state preparation and probing techniques such as internal state manipulation could be useful~\cite{Amico_2018,Scazza_2020}.

We thank D.S. Petrov, F. Scazza, M. Zaccanti, and G. Roati for fruitful discussions. We also thank Z. Hadzibabic, F. Werner, and L. Chambard for comments on the manuscript. This work was supported by the NSF, DARPA, the David and Lucile Packard Foundation, and the Alfred P. Sloan Foundation.  

\newpage
\cleardoublepage
\setcounter{figure}{0}
\setcounter{equation}{0}
\renewcommand{\thefigure}{S\arabic{figure}}

\makeatletter
\def\maketitle{
\@author@finish
\title@column\titleblock@produce
\suppressfloats[t]}
\makeatother

\title{ Supplemental Material for\\ `On the Stability of the Repulsive Fermi Gas with Contact Interactions'}
\maketitle
\onecolumngrid

\section{\texorpdfstring{\RNum{1}}. Box Calibration}

\subsection{A. Volume}
We fit the \emph{in-situ} optical density of atoms in our box with the profile of a conical cylinder of radii $R_1$ and $R_2$, and length $L$. The fitting profile is convoluted with the Airy function to account for the finite imaging and box-projection resolution. The size of a pixel on the OD images (see Fig.~1(b)) is measured to be $2.00(5)~\mu \text{m}$ using the centre-of-mass motion of a free-falling gas. The dimensions of our box are $R_1 = 47(1)~\mu\text{m}$, $R_2 = 44(1)~\mu\text{m}$, and $L  = 111(3)~\mu\text{m}$ (the uncertainties here result from a combination of fitting errors and pixel size uncertainty). Note that the determination of the uncertainty on $V$ takes into account correlations on the uncertainties on $R_1$, $R_2$, and $L$.

\subsection{B. Depth}
\begin{figure}[h!]
    \includegraphics[width=0.6\columnwidth]{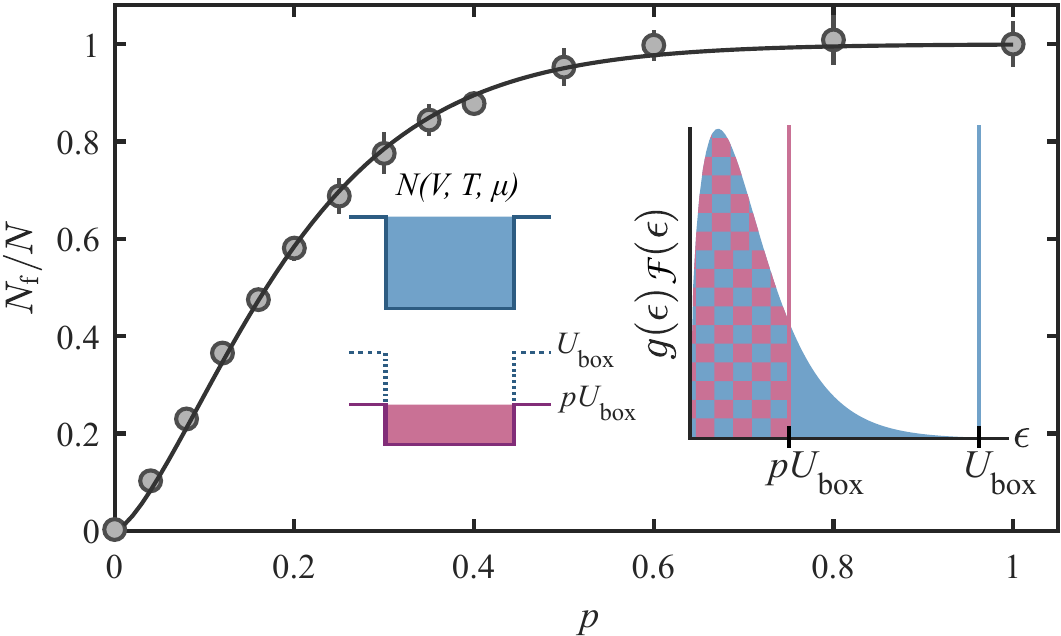}
    \caption{Calibration of the box depth. The black solid line is the fitting using Eq.~(\ref{eq:boxdepth}). The box before and after the truncation is respectively depicted as blue and red, and the areas under $g(\epsilon)\mathcal{F}(\epsilon)$ are shown with the same colors.}
\label{FIG:boxdepth}
\end{figure}

To measure the box depth, we load a weakly repulsive gas (at $a \approx 500~a_0$) into the box at the depth $U_{\mathrm{box}}$ and let the gas thermalize. We first measure the atom number $N$, the temperature $T$, and the chemical potential $\mu$ after the time-of-flight expansion. %after time-of-flight expansion. 

In a second set of experiments, instead of letting the gas expand, we turn off the interactions \emph{in situ}, and change the box depth to $p U_\text{box}$ (where $0 \le p \le 1$). The remaining atoms in the box $N_\text{f}$ give access to the truncated energy distribution (see inset of Fig.~\ref{FIG:boxdepth}). Given the initial $T$ and $\mu$, the ratio $N_\text{f}/N$ is determined by $U_{\mathrm{box}}$ and $p$:
\begin{equation}
\begin{split}
\frac{N_\text{f}}{N} &= \frac{\int_0^{p U_{\mathrm{box}}} \mathrm{d}\epsilon  \;  g(\epsilon)\mathcal{F}(\epsilon)}{\int_0^{U_{\mathrm{box}}} \mathrm{d}\epsilon  \;  g(\epsilon)\mathcal{F}(\epsilon)} \;\stackrel{k_{\text{B}}T, E_{\text{F}} \ll U_{\mathrm{box}}}{\approx} \;\frac{\int_0^{p U_{\mathrm{box}}} \mathrm{d}\epsilon  \;  g(\epsilon)\mathcal{F}(\epsilon)}{\int_0^{\infty} \mathrm{d}\epsilon  \;  g(\epsilon)\mathcal{F}(\epsilon)},\\
\end{split} 
\label{eq:boxdepth}
\end{equation}
where $\mathcal{F}(\epsilon) = 1/(e^{(\epsilon - \mu)/(k_{\text{B}} T)} + 1)$ is the Fermi-Dirac distribution, and $g(\epsilon) = (2m/\hbar)^{3/2} \sqrt{\epsilon}/(2 \pi)^2$ is the density of states per spin state. In Fig.~\ref{FIG:boxdepth}, we show $N_\mathrm{f}/N$ as a function of $p$ for a cloud with $T = 186(12)$~nK and $\mu = k_{\text{B}} \times 41(3)$~nK. From the fit to Eq.~(\ref{eq:boxdepth}), we determine our box depth to be $U_{\mathrm{box}} =  k_{\text{B}} \times 1.6(2) ~\mu\mathrm{K}$.

\section{\texorpdfstring{\RNum{2}}. Preparation of a Metastable Repulsive Fermi Gas}

\subsection{A. Atom Losses During Initial State Preparation}
\begin{figure}[!h]
    \includegraphics[width=0.5\columnwidth]{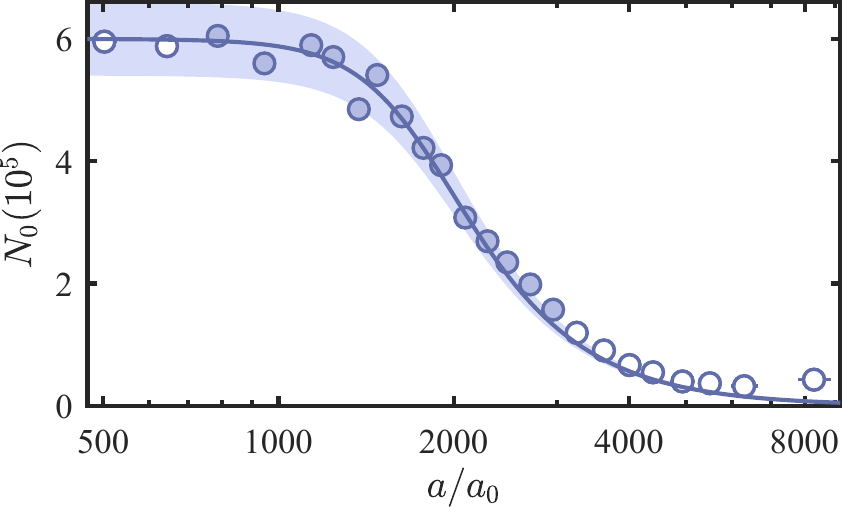}
    \caption{Initial atom number $N_0$ after the interaction ramp and the magnetic field settling time. The point style follows the same criterion as in Fig.~4. The blue line shows the result of the simulation. The shaded band include an error of $10 \%$ on the atom number at the beginning of the ramp, and $20 \%$ on $\epsilon_{\mathrm{kin}}$.} 
\label{fig:N0}
\end{figure}

Here, we use our measurement of $K_3$ to consistently calculate the `initial' atom number $N_0$ from the main text. At any given time, the atom loss follows the equation
\begin{equation}\label{eq:ramp}
\begin{split}
\dot n &= -L_3(t)n^3  \\
& = -3 A \frac{\epsilon_{\text{kin}}}{\hbar}a(B(t))^6 n^3,\\
\end{split}
\end{equation} 
where $a(B)$ is the $s$-wave scattering length as a function of the magnetic field~\cite{zurn_2013}, $B(t) = B_0 + (B_\text{f} -B_0)t/t_{\mathrm{ramp}}$ if $0\leq t \leq t_\text{ramp}$, $B(t)=B_\text{f}$ for $t>t_\text{ramp}$, and $B_\text{f}$ is the final interaction field; $B_0 = 597$~G is the initial field, and $t_{\mathrm{ramp}} = 100~\mathrm{ms}$. Following the notation of the main text, we have  $N_0=N(t_\text{hold}=0)=N(t=t_\text{ramp}+t_\text{settle})$, with $t_\text{settle}=25$~ms. 
 We solve the equation using $N(t=0) = 6 \times 10^5$ atoms at the beginning of the ramp, and we choose a typical $\epsilon_{\text{kin}}/k_{\mathrm{B}} = 500~\mathrm{nK}$.  The simulation agrees very well with the data in the range where $\gamma\approx 3$ (solid points in Fig.~\ref{fig:N0}).

\subsection{B. Competition Between Upper-Branch Decay and Two-body Thermalization}
\begin{figure}[!h]
    \includegraphics[width=0.5\columnwidth]{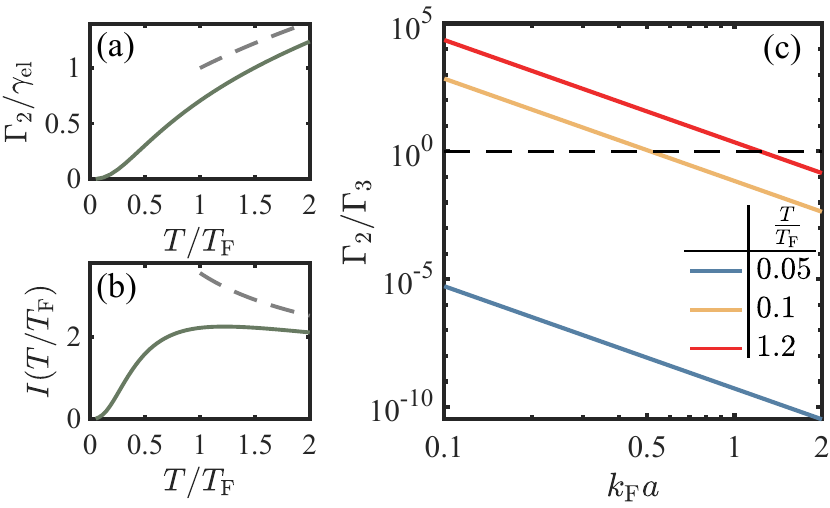}
    \caption{Two-body elastic versus three-body inelastic collisions. (a) Two-body elastic scattering rate, normalized by $\gamma_{\mathrm{el}}$. The solid line shows the result calculated from the Boltzmann equation. The dashed line shows the result in the classical regime, $\Gamma_2/\gamma_{\mathrm{el}} = \sqrt{T/T_{\mathrm{F}}}$. (b) $I(T/T_\mathrm{F})$ versus $T/T_{\mathrm{F}}$. The solid and dashed lines use $\Gamma_2$ from the Boltzmann equation and classical limit approximation respectively. (c) $\Gamma_2/\Gamma_3$ versus $k_{\mathrm{F}} a$ for $T/T_{\mathrm{F}} = 0.05$ (blue), $0.1$ (yellow), $1.2$ (red). The horizontal dashed lines marks $\Gamma_3=\Gamma_2$.}
\label{fig:gamma2}
\end{figure}

Preparing a repulsive Fermi gas in thermal equilibrium requires that two-body elastic collisions dominate over the decay of the `upper' branch. We calculate the two-body elastic collision rate $\Gamma_2$ from the Boltzmann equation assuming the scattering cross section is energy independent, $\sigma = 4\pi a^2$~\cite{luiten_1996,gehm_2003}. Let us consider the process of two atoms with initial energies $\epsilon_1$ and $\epsilon_2$ colliding and having final energies $\epsilon_3$ and $\epsilon_4$. The collision rate is given by
\begin{equation}\label{eq:boltzmann}
\begin{split}
\Gamma_2 =& \frac{m \sigma}{n \pi^2 \hbar^3} \iiiint \mathrm{d}\epsilon_1\mathrm{d}\epsilon_2\mathrm{d}\epsilon_3\mathrm{d}\epsilon_4\; g\big(\mathrm{min}(\epsilon_1,\epsilon_2,\epsilon_3,\epsilon_4)\big) \delta(\epsilon_1+\epsilon_2-\epsilon_3-\epsilon_4)\mathcal{F}(\epsilon_1)\mathcal{F}(\epsilon_2)\big(1-\mathcal{F}(\epsilon_3)\big)\big(1-\mathcal{F}(\epsilon_4)\big).
\end{split}
\end{equation} 
Here, the effects of Pauli blocking of both ingoing and outgoing channels are taken into account. In Fig.~\ref{fig:gamma2}(a), we display $\Gamma_2$ as a function of $T/T_\text{F}$; $\Gamma_2$ is normalized to $\gamma_\text{el}$, which is the classical collision rate $\sigma \langle v_{\mathrm{eff}} \rangle n$ at $T=T_{\text{F}}$ ($\langle v_{\mathrm{eff}} \rangle$ is the thermal-averaged relative velocity between two particles). Two-body elastic scattering expectedly vanishes as $T \rightarrow 0$ due to Pauli blocking on the outgoing channel.

For $\Gamma_3$, we consider the worst case scenario where all three atoms are lost from the repulsive branch per recombination event, i.e. $\Gamma_3 = 3 A \epsilon_{\mathrm{kin}} a^6 n^2/\hbar$. We find that $\Gamma_2/\Gamma_3 = I(T/T_\mathrm{F}) (k_{\mathrm{F}} a)^{-4}$ where $I(T/T_\mathrm{F})$ is a universal function that reaches its maximum at $T/T_{\mathrm{F}} \approx 1.2$, and $I(T/T_\mathrm{F}) \approx 64 \sqrt{2} \pi^{5/2} \sqrt{T_{\mathrm{F}}/T}/(3A)$ in the classical limit (dashed line in Fig.~\ref{fig:gamma2}(b)). Finally we show $\Gamma_2/\Gamma_3$ versus $k_\text{F}a$ for various $T/T_\text{F}$ in Fig.~\ref{fig:gamma2}(c).

\section{\texorpdfstring{\RNum{3}}. The ramp to the imaging field}
\begin{figure}[!hbt]
    \includegraphics[width=0.5\columnwidth]{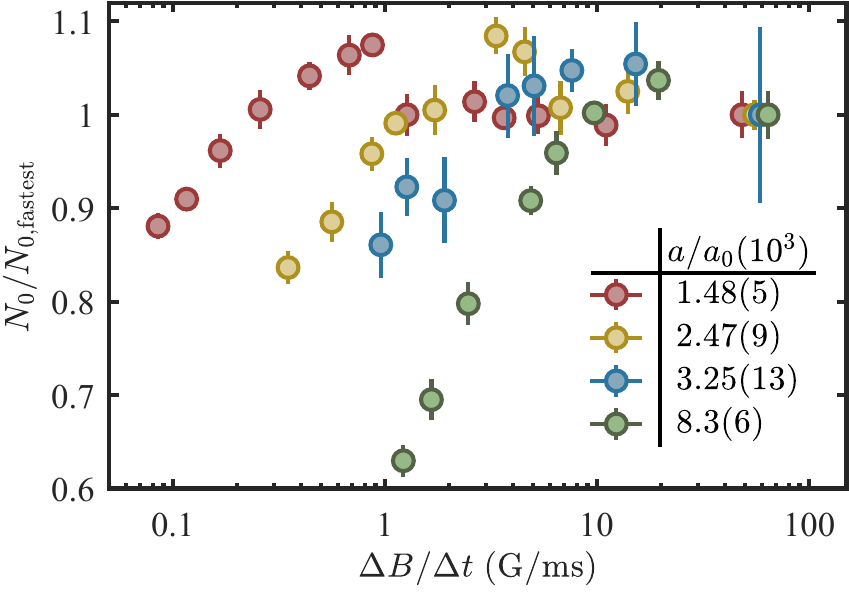}
    \caption{Effect of the ramp speed to the imaging field on the number of atoms detected. The number of atoms is normalized by the value measured by the fastest ramp.}
    \label{fig:rampspeed}
\end{figure}

We verify the robustness of our ramp back to the `imaging' field (near the zero-crossing of $a$) by varying the ramp speed and measuring the detected atom number at $t_\text{hold}=0$. We observe that $N_0$ plateaus for $\Delta B/\Delta t \ge 10 ~\mathrm{G}/\mathrm{ms}$ (Fig.~\ref{fig:rampspeed}). 
The pairs that stay trapped after recombinations are converted to tightly bound molecules during the ramp. We estimate the conversion probability by modeling the ramp as a Landau-Zener sweep where the two discrete levels are the molecular state and the `free atoms at rest' state. The conversion probability is $>99\%$ for the range of $B$ used in the experiment.

\section{\texorpdfstring{\RNum{4}}. Relation between \texorpdfstring{$\omega_3$}{Lg} and \texorpdfstring{$\Omega$}{Lg}}
We integrate the momentum-dependent three-body event rate in a unit volume
\begin{equation}
\omega_3(\mathbf{k}_1,\mathbf{k}_2,\mathbf{k}')= \frac{A}{2} \frac{\hbar a^6}{m} (\mathbf{k}_1-\mathbf{k}_2)^2\\
\end{equation} 
over the phase space density to find $\Omega$, the macroscopic event rate per volume:
\begin{equation}\label{eq:omegatoekin}
\begin{split}
\Omega &=\frac{1}{2}\frac{1}{(2 \pi)^9 }\iiint \mathrm{d}\mathbf{k}_1 \mathrm{d}\mathbf{k}_2 \mathrm{d}\mathbf{k}' \;\omega_3(\mathbf{k}_1,\mathbf{k}_2,\mathbf{k}') \mathcal{F}(\epsilon_{k_1})\mathcal{F}(\epsilon_{k_2})\mathcal{F}(\epsilon_{k'}) \\
&= \frac{1}{2}\frac{A}{2} 
\frac{\hbar a^6}{m} \Big(\frac{1}{(2 \pi)^3 } \int \mathrm{d}\mathbf{k}'\;\mathcal{F}(\epsilon_{k'}) \Big) \Big(\frac{1}{(2 \pi)^6 } \iint \mathrm{d}\mathbf{k}_1 \mathrm{d}\mathbf{k}_2 \; (\mathbf{k}_1-\mathbf{k}_2)^2 \mathcal{F}(\epsilon_{k_1})\mathcal{F}(\epsilon_{k_2})\Big),\\
\end{split}
\end{equation} 
where $\epsilon_{k} = \hbar^2 k^2/(2m)$. The factor $1/2$ accounts for the indistinguishability of the identical atoms. As
\begin{equation}
\begin{split}
n &= \frac{1}{(2 \pi)^3 } \int \mathrm{d}\mathbf{k} \;\mathcal{F}(\epsilon_{k}) \;\; \text{and} \;\; \epsilon_{\mathrm{kin}} =\frac{1}{n}\frac{1}{(2 \pi)^3 }\int \mathrm{d}\mathbf{k} \; \frac{\hbar^2 k^2}{2m}\mathcal{F}(\epsilon_{k}),\\
\end{split}
\end{equation} 
we find
\begin{equation}
\begin{split}
\Omega &= A\frac{a^6}{\hbar} \epsilon_{\mathrm{kin}}n^3.
\end{split}
\end{equation} 
The recombination coefficient $K_3$ therefore scales linearly with the average kinetic energy per particle:
\begin{equation}\label{eq:L3}
\begin{split}
K_3 = \frac{\Omega}{n^3} =  A\frac{a^6}{\hbar} \epsilon_{\mathrm{kin}}.
\end{split}
\end{equation} 

\section{\texorpdfstring{\RNum{5}}. Energy and Temperature Dynamics under Three-body Recombination}
Since $\Omega\propto \epsilon_\text{kin}$, the average kinetic energy per particle should monotonously decrease as a result of three-body loss alone.
Indeed, the evolution of the energy density $\varepsilon= n \epsilon_{\mathrm{kin}}$ follows
\begin{equation}\label{eq:Ekevo}
\begin{split}
\dot\varepsilon &= -\frac{1}{2}\frac{1}{(2 \pi)^9 } \iiint \mathrm{d}\mathbf{k}_1 \mathrm{d}\mathbf{k}_2 \mathrm{d}\mathbf{k}'\; (\epsilon_{k_1} + \epsilon_{k_2} + \epsilon_{k'})\omega_3(\mathbf{k}_1,\mathbf{k}_2,\mathbf{k}')\mathcal{F}(\epsilon_{k_1})\mathcal{F}(\epsilon_{k_2})\mathcal{F}(\epsilon_{k'}) \\
&= -\frac{1}{2} \frac{A}{2} 
\frac{\hbar a^6}{m} \frac{1}{(2 \pi)^9 }\Big(\int \mathrm{d}\mathbf{k}'\;\frac{\hbar^2 k'^2}{2m}\mathcal{F}(\epsilon_{k'}) \iint \mathrm{d}\mathbf{k}_1 \mathrm{d}\mathbf{k}_2 \; (\mathbf{k}_1-\mathbf{k}_2)^2 \mathcal{F}(\epsilon_{k_1})\mathcal{F}(\epsilon_{k_2})\\
&+ \int \mathrm{d}\mathbf{k}'\;\mathcal{F}(\epsilon_{k'}) \iint \mathrm{d}\mathbf{k}_1 \mathrm{d}\mathbf{k}_2 \; \frac{\hbar^2 (k_1^2+k_2^2)}{2m}(\mathbf{k}_1-\mathbf{k}_2)^2 \mathcal{F}(\epsilon_{k_1})\mathcal{F}(\epsilon_{k_2})\Big)\\
&= -2 A\frac{a^6}{\hbar} \varepsilon^2 n \Big(1 +  \frac{5}{6}\frac{\ell_7 \ell_3}{\ell_5^2}\Big),
\end{split}
\end{equation}
where $\ell_s \equiv \mathrm{Li}_{s/2} (-e^{\mu/k_{\mathrm{B}}T})$, and $\mathrm{Li}_{s/2}$ is the polylogarithm function of order $s/2$. The quantum degeneracy is given by $T/T_{\mathrm{F}} = (-3 \sqrt{\pi}\ell_3/4)^{-2/3}$. The evolution of the average kinetic energy per particle under the three-body-loss rarefaction is governed by the coefficient $\theta_\epsilon$:
\begin{equation}
\begin{split}
\theta_{\epsilon}&= \frac{\mathrm{d}\epsilon_{\mathrm{kin}}/\epsilon_{\mathrm{kin}}}{\mathrm{d}N/N} = \frac{\dot \epsilon_{\mathrm{kin}}/\epsilon_{\mathrm{kin}}} {\dot n/n} = \frac{\dot \varepsilon/\varepsilon} {\dot n/n} - 1\\
&=\frac{5}{9} \frac{\ell_7 \ell_3}{\ell_5^2} - \frac{1}{3}.
\end{split}
\end{equation}
We find that $\theta_{\epsilon} > 0$, so that $\epsilon_{\mathrm{kin}}$ always decreases. We define an analogous coefficient for the evolution of $T$ under the same process:
\begin{equation}
\begin{split}
\theta &= \frac{N}{T}\left(\frac{\partial T}{\partial N}\right)_V = \frac{2}{9} \Big(\frac{9\ell_3^2 -6\ell_5 \ell_1}{{3\ell_3^2 - 5 \ell_5 \ell_1}}-\frac{5\ell_7\ell_3 \ell_1}{{3\ell_5 \ell_3^2 - 5 \ell_5^2 \ell_1}} \Big).
\end{split}
\end{equation}
In contrast to $\theta_{\epsilon}$, $\theta$ changes sign at $(T/T_\text{F})^* \approx 0.71$.

\section{\texorpdfstring{\RNum{6}}. Average kinetic energies of the recombination products}
We calculate the kinetic energies of the recombination products, averaged over the phase space density and recombination probability. By energy and momentum conservation, the final kinetic energies of the molecule $\epsilon_\text{m}$ and the free atom $\epsilon_\text{a}$ are given by 
\begin{equation}
\begin{split}
\epsilon_{\text{m}} &= \frac{1}{3} \epsilon_{\text{b}} + \frac{5}{9}\frac{\hbar^2}{2m} (k_1^2 + k_2^2 + k'^2) - \frac{2}{9}\frac{\hbar^2}{2m} (\mathbf{k}_1\cdot\mathbf{k}_2 + \mathbf{k}'\cdot\mathbf{k}_2 + \mathbf{k}'\cdot\mathbf{k}_1)\\
\epsilon_{\text{a}} &= \frac{2}{3} \epsilon_{\text{b}} + \frac{4}{9}\frac{\hbar^2}{2m} (k_1^2 + k_2^2 + k'^2) + \frac{2}{9}\frac{\hbar^2}{2m} (\mathbf{k}_1\cdot\mathbf{k}_2 + \mathbf{k}'\cdot\mathbf{k}_2 + \mathbf{k}'\cdot\mathbf{k}_1).
\end{split}
\end{equation}
The averaged quantities are 
\begin{equation}
\begin{split}
\langle\epsilon_{\text{m/a}}\rangle &= \frac{1}{\Omega}\frac{1}{2}\frac{1}{(2 \pi)^9 }\iiint \mathrm{d}\mathbf{k}_1 \mathrm{d}\mathbf{k}_2 \mathrm{d}\mathbf{k}' \;\epsilon_{\text{m/a}}\omega_3(\mathbf{k}_1,\mathbf{k}_2,\mathbf{k}') \mathcal{F}(\epsilon_{k_1})\mathcal{F}(\epsilon_{k_2})\mathcal{F}(\epsilon_{k'}). \\
\end{split}
\end{equation} 
In the limit $T\rightarrow 0$, we find
\begin{equation}
\begin{split}
\langle\epsilon_{\text{m}}\rangle &= \frac{\epsilon_{\text{b}}}{3} + \frac{254}{315}E_{\text{F}}, \;\;\langle\epsilon_{\text{a}}\rangle = \frac{2\epsilon_{\text{b}}}{3} + \frac{349}{315}E_{\text{F}}.
\end{split}
\end{equation}
The grey bands in Fig.~4 are determined accordingly.

\section{\texorpdfstring{\RNum{7}}. Effect of \texorpdfstring{$\epsilon_{\text{kin}}(t)$}{Lg} on \texorpdfstring{$L_3$}{Lg}}
\begin{figure}[!bth]
    \includegraphics[width=0.9\columnwidth]{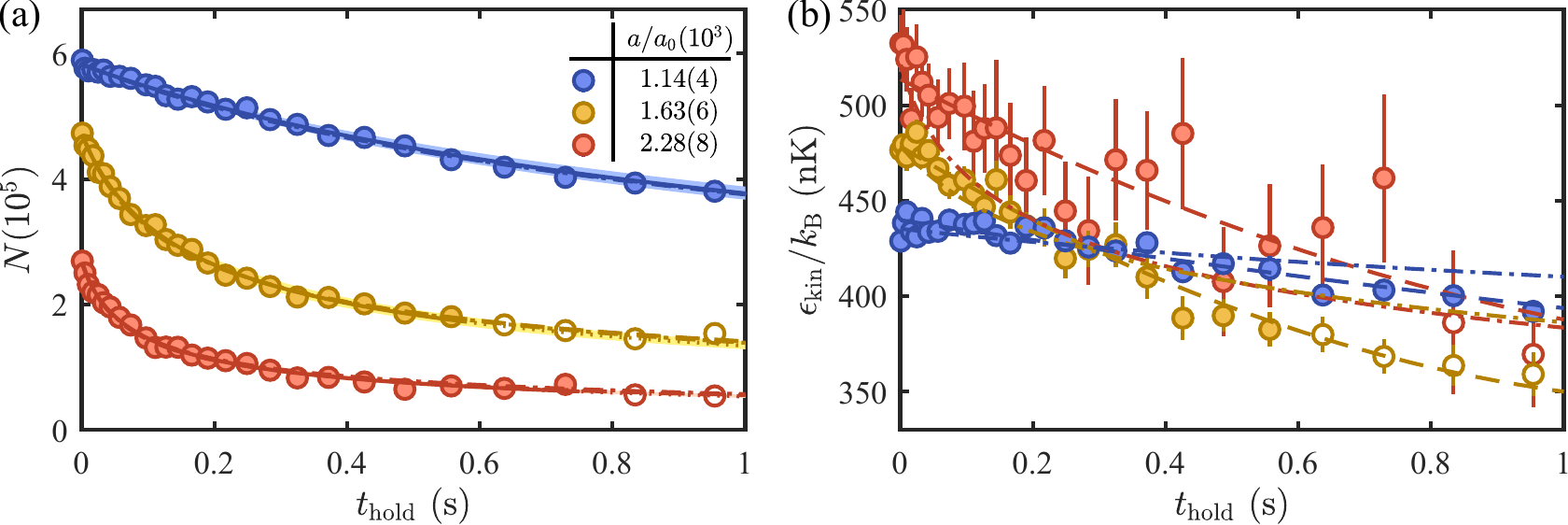}
    \caption{Effect of the variation of $\epsilon_\text{kin}(t)$ on the determination of $L_3$. (a) Dynamics of $N(t)$. The data points and solid-to-dotted lines are the same as in Fig.~2 of the main text (i.e. assuming $L_3$ is constant with $t_\text{hold}$). The bands are $L_3$-constant $N(t)$ with $20\%$ total error on $\epsilon_\text{kin}$. The dashed lines are numerical calculations including the evolution of $\epsilon_\text{kin}$, see panel (b). The dashed-dotted lines are the direct theoretical predictions (same in (b)). (b) Dynamics of $\epsilon_\text{kin}(t)$. The dashed lines are smooth interpolations used to compute the dashed lines in panel (a).}
\label{fig:l3t}
\end{figure}

In the main text, we extracted $L_3$ assuming it is constant over time. However, we established that it is linearly dependent on $\epsilon_{\mathrm{kin}}$, and that $\epsilon_{\mathrm{kin}}$ monotonously decreases during three-body losses. Here we discuss this apparent inconsistency by including the evolution of $\epsilon_\text{kin}$ in the dynamics of $N(t)$ and comparing the results to the $L_3$-constant fits. In Fig.~\ref{fig:l3t}, we show the original fits of the main text as solid-to-dotted lines and the bands are $L_3$-constant $N(t)$ considering the limiting case of $\pm 10\%$ error on $\epsilon_{\text{kin}}$. The dashed lines in Fig.~\ref{fig:l3t}(a) are the solutions $N(t)$ taking into account the experimentally-measured $\epsilon_\text{kin}(t)$ (using the smooth interpolations for $\epsilon_\text{kin}(t)$, dashed lines in Fig.~\ref{fig:l3t}(b)). For reference, the dashed-dotted lines in Fig.~\ref{fig:l3t} are the solutions using the coupled equations for $\varepsilon(t)$ and $n(t)$.

As seen in Fig.~\ref{fig:l3t}(a), simulations of $N(t)$ including $\epsilon_{\text{kin}}(t)$ lie well within the bands and show little difference from $L_3$-constant fits. The variation of $\epsilon_\text{kin}(t)$ is included in the determination of the error bars on $L_3/\bar\epsilon_{\text{kin}}$ in Fig.~4(b).


\begin{thebibliography}{72}
\bibitem [{ima()}]{imaging}%
  \BibitemOpen
  \href@noop {} {}\bibinfo {note} {$\epsilon_\text{b} = \hbar\Gamma$, where
  $\Gamma$ is the natural linewidth of the imaging resonance, corresponds to $a
  \approx 300~a_0$.}\BibitemShut {Stop}%
\bibitem [{one()}]{onebody}%
  \BibitemOpen
  \href@noop {} {}\bibinfo {note} {Including the effect of one-body losses in
  the equation for $\dot{n}$ yields: $\dot{n} = -L_3 n^3-n/\tau_1$, where
  $\tau_1=32(2)$~s is the background-gas limited lifetime. This effect is small
  for all but the smallest $a$ we explored.}\BibitemShut {Stop}%
\bibitem [{eps()}]{epsilonkin}%
  \BibitemOpen
  \href@noop {} {}\bibinfo {note} {The x-axis values in Fig.3(a) are the
  kinetic energies averaged over the range $L_3$ is
  extracted~\cite{SuppMat}.}\BibitemShut {Stop}%
\bibitem [{Aun()}]{Auniv}%
  \BibitemOpen
  \href@noop {} {}\bibinfo {note} {More precisely, $A$ is predicted to be a
  universal function of the mass ratio $m'/m$ (where $m'$ is the third
  participant's mass) provided $m'/m<13.6$~\cite{Petrov_2003}.}\BibitemShut
  {Stop}%
\bibitem [{dep()}]{deposit}%
  \BibitemOpen
  \href@noop {} {}\bibinfo {note} {In retrospect, it is puzzling that the
  effect of the deposited energy of produced molecules is not seen, given that
  the atom-molecule scattering length $\approx
  a$~\cite{Skorniakov_1957}.}\BibitemShut {Stop}%
\bibitem [{uni()}]{unitary}%
  \BibitemOpen
  \href@noop {} {}\bibinfo {note} {One can naively estimate the unitary limit
  of $L_3$ by using the prescription $a\rightarrow\lambda_T = h/\sqrt{2 \pi m
  k_{\text{B}} T}$. For $T \approx 500~\mathrm{nK}$, we find
  $L_3/\epsilon_{\text{kin}} \approx 6\times 10^{-20}~\text{cm}^6/(\text{s}
  \cdot \mathrm{nK} \cdot k_{\text{B}})$, well above our highest
  $L_3/\bar\epsilon_{\text{kin}}$.}\BibitemShut {Stop}%
\bibitem [{gam()}]{gamma3}%
  \BibitemOpen
  \href@noop {} {}\bibinfo {note} {Depending on whether the third atom is
  trapped after a three-recombination event, $\Gamma_3/(K_3 n^2)$ is a number
  between 2 and 3. Here we consider the worst case scenario.}\BibitemShut
  {Stop}%
\bibitem [{kFa()}]{kFalim}%
  \BibitemOpen
  \href@noop {} {}\bibinfo {note} {Note that this limit is only marginally
  consistent, since it is calculated assuming the universal $K_3$ law; this
  assumption is expected to break down for $\epsilon_\text{b}\lesssim
  E_\text{F}$.}\BibitemShut {Stop}%
\end{thebibliography}
\end{document}